\documentclass[aps,preprint, floatfix, superscriptaddress,]{revtex4}

\pdfoutput=1

\usepackage{amsmath}
\usepackage{amssymb} \usepackage{dcolumn}
\usepackage{epsfig}
\usepackage{graphics}
\usepackage{graphicx}
\usepackage{slashed,epsfig}
\usepackage{longtable}
\usepackage{color}

\definecolor{darkgreen}{rgb}{0,0.5,0}
\definecolor{purple}{rgb}{0.5,0,0.5}
\definecolor{nblue}{rgb}{0.0,0.0,0.50}
\definecolor{scarlet}{rgb}{1.0,0.2,0}
\usepackage[colorlinks=true, pdfstartview=FitV, linkcolor=purple, citecolor= blue, urlcolor=blue]{hyperref}

\newcommand{\beq} {\begin{equation}}
\newcommand{\eeq} {\end{equation}}
\newcommand{\beqa} {\begin{eqnarray}}
\newcommand{\eeqa} {\end{eqnarray}}

\begin{document}

\bigskip{}

\title{Dirac Quantization Condition Holds with Nonzero Photon Mass }

\author{Alfred Scharff Goldhaber}
\affiliation{C.N. Yang Institute for Theoretical Physics, State University of New York, Stony Brook, NY 11794-3840, USA}
\email{goldhab@max2.physics.sunysb.edu}
\author{Ricardo Heras}
\affiliation{Department of Physics and Astronomy,
 University College London, London, WC1E 6BT, UK}
 \email{ricardo.heras.13@ucl.ac.uk}
\begin{abstract}
{ \indent \ \ }  Dirac in 1931 gave a beautiful argument for the quantization of electric charge, which required only the existence in the universe of one magnetic monopole,  because gauge invariance of the interaction between the pole and any charge could hold only if the product of the charge and the pole strength were quantized in half-integer multiples of the reduced Planck constant.  However, if the photon had a nonzero mass, implying exponential decrease of the flux out of an electric charge, then Dirac's argument might seem to fail. We demonstrate that the result still should hold.  The key point is that magnetic charge, unlike electric charge, cannot be screened, so that  on any surface enclosing the pole Dirac's string, or equally  the Wu-Yang gauge shift, must be present, and to make either of these invisible to charged particles the quantization condition is required.\end{abstract}


\maketitle

\date{\today}

\section{Introduction}

Dirac's 1931 paper \cite{1}, best-known for giving only the second prediction of a new particle, the anti-electron or positron  (the first being Einstein's introduction of the quantum of light, the photon \cite{2}), was devoted largely to consideration of a hypothetical particle, a magnetic monopole, still undiscovered today.  By examining the interaction between such a pole and an ordinary electric charge, he deduced that the product of the electric and magnetic charges must be quantized in integer multiples of the smallest quantum of angular momentum, $\hbar/2$. There is a dramatic implication:  the existence of even one monopole in the universe would imply the quantization of all electric charges!  

Later authors \cite{3,4} noted that Dirac's result could be obtained from quantization of the angular momentum in the electromagnetic field resulting from the crossed electric and magnetic fields of charge and pole.   However, if the photon had a nonzero mass, the exponential falloff of the static electric field from the charged particle would invalidate that second form of the argument:  the electromagnetic angular momentum would decline in magnitude with increasing distance between charge and pole, so that its magnitude could not have a fixed, quantized value.  This led to a number of papers \cite{5,6} saying that nonzero photon mass is incompatible with Dirac's quantization condition. Here we show that even though the angular-momentum quantization argument  fails under these circumstances, Dirac's gauge-invariance considerations continue to hold, and therefore the quantization condition remains valid. Our analysis makes use of the `modern' method of introducing photon mass, electromagnetic coupling to an electrically charged scalar field, introduced independently by Higgs \cite{7}, Englert and Brout \cite{8}, and Hagen, Guralnik, and Kibble \cite{9} in the early 1960s. As we shall see, if the photon mass be introduced in the `old-fashioned' Proca way, as a fixed constant, then gauge invariance manifestly is violated, but nevertheless the Dirac argument still holds.

\section{The Dirac argument}

In electrodynamics potential functions are introduced, a scalar potential $V$ and a vector potential ${\bf A}$, with the electric field taking the form
 ${\bf E}=-\nabla V-\partial {\bf A}/\partial t$, and the magnetic field
  ${\bf B} = \nabla \times {\bf A}.$ Because of the well-known fact that the divergence of the curl of a vector field must vanish, one might think that this would preclude the possibility of magnetic monopoles, but Dirac easily overrode this concern, describing the pole as one end of a ``fictitious'' singular string of magnetic flux, where the magnitude of the flux would be just such as to render it invisible to charged particles. With this ingenious description of the pole, Dirac derived his quantization condition. In creating his apparently singular description, Dirac invented the concept of `fiber bundles' about five years before mathematicians discovered it \cite{10}, of course in complete ignorance of Dirac's work.

A less singular way to describe Dirac's argument, along the lines of modern fiber bundle theory,  was presented by Wu and Yang \cite{11}, who introduced two singular vector potentials, one with its singular line in the direction of the south pole of a sphere centered on the monopole, whose potential is used for the northern hemisphere centered on the pole, while the second vector potential has its singularity in the direction of the north pole, and is used to describe the magnetic field in the southern hemisphere. In the region around the equator where both potentials are nonsingular, there is a gauge transformation connecting them, {\it provided} the product of electric and magnetic charges is quantized as described earlier.  The transformation function is $e^{in\phi}$, where $\phi$ is the azimuthal angle about the polar direction,   and $n$ is twice the number of quantum units in the product of electric and magnetic charges.  Only if the function is continuous on the circle, i.e., only if $n$ is an integer, is this an acceptable gauge transformation, and hence only in that case is the theory well defined.  For Dirac, with his singular string, the quantization condition is necessary to make the string invisible.  This means that a charged-particle wave diffracted around the string would go forward exactly as if no string were present.

\section{Giving mass to the photon}

As envisioned by de Broglie and most fully articulated by Proca \cite{12}, a mass $\mu$ is provided to the electromagnetic field by adding to the Lagrangian a term proportional to $\mu^2({\bf A}^2-V^2)$.  This yields terms proportional to $\mu^2$ in the equations of motion, which mean that for a free photon the relation between energy and momentum is given by $\hbar\omega=\sqrt{(\hbar c{\bf k})^2+(\hbar\mu c)^2}$, where $\omega$ is the circular frequency of the photon,  ${\bf k}$ is its wave vector, and $\mu$ is the mass expressed in units of inverse length.  A trouble with this formulation is that it breaks the gauge invariance of electrodynamics, imposing the Lorenz gauge on the electromagnetic potential functions.  With time it has become clearer and clearer that gauge invariance is an important principle, and in fact where it is relevant that invariance is necessary for consistency.
The violation of gauge invariance implied by a constant photon mass goes beyond the imposition of the Lorenz gauge: the 4-vector potential now acts as a `current' that generates additional observable electromagnetic fields (Proca's four-potential involves an additional degree of freedom \cite{13}).  

The modern way to maintain gauge invariance, while accommodating a nonzero photon mass, is through a mechanism introduced by Higgs \cite{7}, Englert and Brout \cite{8}, and Guralnik, Hagen, and Kibble \cite {9}, independently in the early 1960's. The Higgs mechanism is one in which gauge transformations act multiplicatively on a field carrying electric charge, or a set of charges.  For the Abelian case (which is what interests us here), this mechanism is equivalent to the (additive) St\"{u}ckelberg mechanism \cite{14}, in which the gradient of a scalar field is added to the vector potential (the St\"{u}ckelberg model can be considered as the free Abelian Higgs model \cite{15}).  What we shall see below, however, is that even the old-fashioned constant mass term,  $\mu^2({\bf A}^2-V^2)$, allows the Wu-Yang mapping to proceed, and hence leaves the theory consistent.

A new, electrically charged  scalar field is introduced, whose action includes an energy density which is minimized for a nonzero value of the field magnitude (the vacuum expectation value or ``vev" of the field).  For the covariant derivative of this Higgs field to vanish in the presence of some background gauge field configuration, the four-vector potential must be the four-gradient of the phase $\alpha$ of a scalar phase factor $e^{i\alpha}$.  Thus we see that  the photon mass is given by the vev of the Higgs field multiplied by the electric charge of that field: Here the point is that if the vector potential is the four-gradient of a scalar function, $A_\mu=\partial_\mu\Phi$, then the associated field strength $F_{\mu\nu}$ vanishes, so that this is a trivial gauge field, and  should be unobservable.  For that to be correct, the covariant derivative of the Higgs field $H$  [i.e., $(\partial_\mu -iqA_\mu)H$] must vanish, which will be true if the spatial dependence of $H$ is given by $H=H_0e^{iq\Phi}$, and of course $H_0$ is the vev of $H$.

The important difference between the Proca mass mechanism and the Higgs mass mechanism is that the Higgs field is a dynamical field, and as a result affects the electric field of a charge and the magnetic field of a pole in different ways.  For the electric field, one has exactly the same exponential fall-off of the electric field for either mechanism.  The Higgs field screens the electric charge, so that in an arbitrarily  large sphere centered on a charged particle the total charge is zero.  This has been called the `local charge' which is measured by the electric flux coming out of the region.  On the other hand, in response to an applied magnetic field, the particle wave function suffers a relative phase shift, when taking two different paths, equal to the product of the charge and the enclosed magnetic flux between the two paths.\cite{16}  The charge thus measured has been called the `Aharonov-Bohm charge' \cite{17} which does {\it not} vanish even though the local charge is zero.  The reason is that as the particle moves, the screening charge simply rearranges itself, with no net flow of current, so that only the particle is sensitive to the magnetic vector potential.

We now are ready to discuss how the Higgs field would affect a magnetic monopole.  Because the Higgs field does not carry magnetic charge, it cannot screen a pole.   However, as Dirac showed in his original paper, a charged-particle wave function, of which the Higgs field is an example, must have a line of zeros emanating from the pole.  Because in the vicinity of that line the field departs from its vev, there must be an energy diverging linearly with radius associated with this configuration.  The only way to truncate this divergence would be to find an antipole at some finite distance from the pole, so that the line of zeros would be finite in extent.  The magnetic flux emanating from the monopole might still be spherically symmetric. If so, the magnetic field of the pole is not screened, but magnetic charge is \it {confined}\rm, because there must always be an anti-pole some finite distance from the pole. The argument just given, that monopoles are confined in the presence of
a Higgs field  because they have lines of zeros of the Higgs field coming out of them, was first presented by Nielsen and Olesen \cite{18}, who used this as a `dual model' for `electric' confinement of quarks. Note that an immediate consequence of the above discussion is that the product of the magnetic charge with the electric charge of the Higgs field must be quantized.  This leaves open the question of whether the charge of any other particle must be quantized.

Under the circumstances just described, the magnetic field of the pole is  the same as it would have been if the Higgs field were not present, so of course the quantization condition should hold exactly as before. However, there is another possibility, first discussed by Abrikosov \cite{19} for magnetic fields interacting with superconductors: the Higgs field, as a macroscopic field with an electric charge coupling, includes as an example  the wave function of a superconductor. Depending on the details of the dynamics, the line of zeros of the Higgs field can also be the center of a tube carrying the total magnetic flux emanating from the pole.
In this case, Dirac's argument still applies:  choosing the line of zeros of the Higgs field to be the $z$ axis, we let the vector potential take the form ${\bf A} = B{\hat{z}}\times {\bf r}/2,$ with $B=4g/R^2.$  Outside the radius $R$ of the tube, we may set ${\bf A}=0$, which clearly assures that its curl, and hence $\bf B$, vanishes.  The gauge transformation on a charged-particle wave function at that boundary then is given by  $e^{i\alpha}=e^{iq\Phi\phi/2\pi}$, which requires for consistency $q\Phi=2\pi n $ (with $q$ the charge of the particle), precisely the Dirac quantization condition! (note  $\Phi=4\pi g$).

We see that, regardless of the form of the distribution of magnetic flux out of the pole, we get the same quantization condition. There is an important point to note here:  In the presence of screening, the net electric flux out of a charge goes to zero exponentially with distance from the charge.  Thus, if we want to measure the charge by its long-range electric field, that charge is zero.  Stated differently, because the polarization of the medium completely compensates the charge of the particle, the `local charge' of the particle is zero.   On the other hand, if we imagine carrying the charge around a region of magnetic flux. its wave function should acquire a phase proportional to the product of charge times flux.  This phase would be observable unless that product were quantized.  So to make the flux invisible, it must take a quantized value, and that means the product of charge and pole strength must be quantized.   The phase in question is called the Aharonov-Bohm phase \cite{19}, after their proposal that, for arbitrary value of the flux, there would be an observable effect from interference between  two parts of a wave, one  passing to the right of the flux and the other passing to the left of the flux.   Note that this AB phase is unaffected by the presence of electrical screening in the medium through which the charge passes, and hence is a measure of what may be called the `Aharonov-Bohm' charge of the particle, in other words the charge at the core, unaffected by the screening supplied by the medium.  Thus, even if the local charge measured by the net electric flux out of the particle vanishes because of screening, the AB charge is unaffected, and continues to be observable.

After all this, we should return to the original Proca mass term, and ask if it really would give different results.  The Proca mass also can only attenuate electric fields by, effectively, supplying screening charge.  It cannot attenuate the magnetic field of a monopole, because even with the Proca term there is no divergence of magnetic fields, and hence no possibility of screening magnetic pole strength.   Expressed differently, the Proca term (or for that matter the Higgs coupling) {\it could} attenuate the circulating magnetic field around a wire, but not the radial field of a monopole. This makes it tempting to say that the gauge-invariance-violating Proca mass would give the same result as the Higgs field, i.e., the Dirac condition.  Let's investigate this.

Imagine once again that the magnetic flux of the monopole were restricted to a tube coming out of the pole.  In the absence of the charged Higgs field, a constant photon mass $\mu$ would produce exponential decay of the field with radius about the center of the tube.  However, because the divergence of the magnetic field vanishes, the total magnetic flux must be the same through a sphere of  any radius as for the massless case.   Thus, to have an everywhere non-singular vector potential one must again introduce the Wu-Yang construction \cite{11}, and  the Wu-Yang gauge transformation can only be a pure gauge transformation if the product of the magnetic charge and any electric charge obeys $4\pi qg=2n\pi$.   This conclusion is independent of whether the magnetic flux is isotropic about the monopole, or concentrated in a tube originating at the pole.

Thus, the conclusions we obtained by thinking in terms of a Higgs field work just as well for a Proca mass.   The primacy of gauge invariance  in modern theoretical physics is strikingly verified by these considerations, and in particular insures the Dirac quantization condition, with or without nonzero photon mass.  To summarize, no matter how the flux out of the monopole is distributed, its total value through any closed surface centered on the monopole is $4\pi g$.  For the vector potential to be nonsingular on that surface, there must be a closed curve on the surface across which the Wu-Yang gauge transformation is implemented.  This can only be a gauge transformation on all charged-particle wave functions if the Dirac quantization condition holds.  Of course, though Dirac's string construction is singular, it also gives a valid description of the gauge field, and the string also must extend as far as the flux, so that formally there is no net flux through any surface surrounding the pole.

This work is dedicated to Professor C.N. Yang on his 95th birthday anniversary.

{}
\end{document}